# A Unified and Economical Approach to Teaching Higher Secondary Electricity Experiments

Sanjoy Kumar Pal[1,3*], Papun Mondal[1], Pradipta Panchadhyayee[2,3], Anirban Samanta[4], and Subhash Chandra Samanta[4]

[1]Anandapur High School, Anandapur, Paschim Medinipur, West Bengal, India
[2]Department of Physics (UG & PG), Prabhat Kumar College, Contai, Purba Medinipur, India
[3]Institute of Astronomy, Space and Earth Science, Kolkata -700054, W. B., India
[4]IAPT-Midnapore College Centre for Scientific Culture, Paschim Midnapore, W. B., India
[*]E-mail: sanjoypal83@gmail.com

## Abstract

In both rural and urban educational settings, science education is often hindered by limited access to lab resources and intimidating, complex instruments. This paper introduces a low-cost, homemade experimental apparatus—built using a mobile charger, nichrome wire, galvanometer, and digital multimeter—that enables educators to perform key higher secondary electricity experiments. The Indigenous Metre Bridge (IMB) has proven to be an intuitive, user-friendly tool that not only bridges theoretical and practical learning but also reduces student apprehension toward lab work. Its simplicity and accessibility exemplify how frugal innovation can transform physics education.

## Introduction

In both rural and urban Indian schools, practical science education often falls short—rural areas face equipment shortages, while urban students frequently avoid labs due to fear of complex instruments. As a high school teacher in rural Bengal, I have seen these challenges firsthand. To overcome them, we have developed a simple, ultra-low-cost Indigenous Metre Bridge (IMB) using everyday materials. Designed for ease, safety, and accessibility, the IMB encourages hands-on learning and makes experimental physics approachable for all students, regardless of their background or prior experience [1-4].

At the heart of this system has been a nearly one-meter length of nichrome wire, stretched tightly between two terminal bolts fixed on a wooden base. Centimetre graph paper is struck on wooden base for measurement of length.  Four separate connecting wires—two on each side—have been secured using the same screws on either end. On the left side (A side), we have attached two red-covered wires, and on the right side (B side), two black-covered wires. One of the red wires on the left has been connected to the positive terminal of a mobile charger, while one of the black wires on the right has been connected to the charger's negative terminal. This arrangement has formed the experimental bridge wire, functioning similarly to that in a standard metre bridge or potentiometer setup. The mobile charger has served as a simple and innovative substitute for a traditional power supply. To complete the circuit, we have used a galvanometer—functioning both as a deflection indicator and a current detector—and a digital multimeter (DMM) for accurate measurements of voltage, resistance, and current. This minimal yet effective set of tools has allowed students to perform a variety of standard electricity experiments with ease and confidence.

The following sections have described, in detail, how this setup has enabled students to perform a series of standard experiments, each having contributed significantly to their conceptual and practical understanding of electricity and electronics.

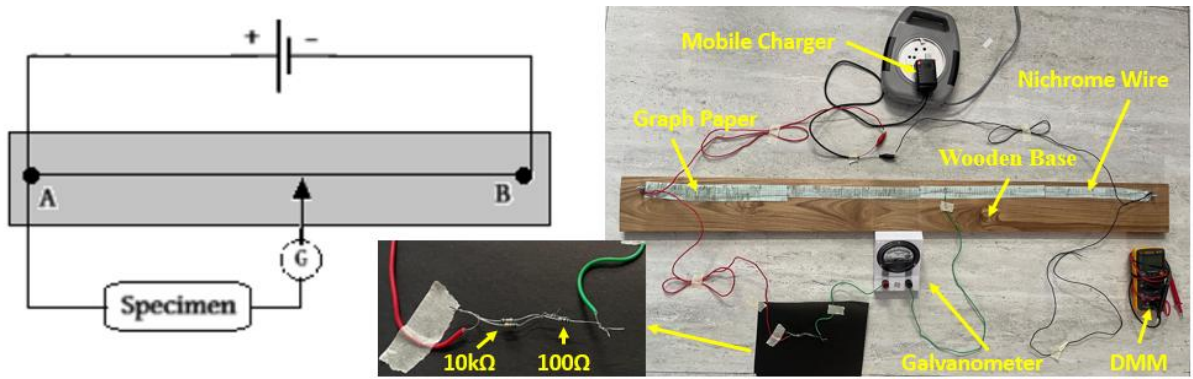

Fig 1: Indigenous Metre Bridge (IMB) circuit diagram (Left), set-up (Right)

## Experiment 1: Determination of EMF and Internal Resistance of a Mobile Charger

We have investigated the internal characteristics of a mobile charger through a hands-on experiment. Using a digital multimeter (DMM) in voltage mode, we have measured the open-circuit voltage (EMF) of the charger to be 5.276 V. We have then connected the charger to an iron-manganese-brass (IMB) wire and have measured the voltage drop across it under load, which has come out to be 3.306 V. The resistance of the wire has been found to be 5.4 Ω. B applying the relation $E = \frac{(E-V)R}{V}$, we have determined the internal resistance of the charger to be about 3.22 Ω. This activity has helped students understand the non-ideal nature of power supplies and has provided a meaningful context for applying theoretical circuit analysis in a practical setting.

## Experiment 2: Calibration and Resistance Measurement of a Galvanometer

In this experiment, we have connected a 100-ohm carbon resistor in series with a galvanometer, an additional 10 kΩ resistor, and a segment of IMB wire to limit the current and protect the galvanometer. By varying the position of the contact on the wire, we have adjusted the current until we have obtained a measurable deflection of d=10 divisions on the galvanometer scale. We have used a digital multimeter (DMM) to measure the voltage drop across the 100-ohm resistor, which has been 17 millivolts. From this, we have calculated the current through the galvanometer using Ohm's law: $I_G = \frac{V}{R}$

We have then calculated the figure of merit (FOM) of the galvanometer—defined as the current per division of deflection—as: $FOM = \frac{I_G}{d}$ = 17μA/div.
With the measured voltage across the galvanometer $V_G$ = 8.67 millivolts, we have calculated its internal resistance of galvanometer is $R_G = \frac{V_G}{I_G}$ = 51 Ω.

This experiment has helped us appreciate the sensitivity and limitations of analog measuring instruments and the importance of incorporating external resistance to prevent damage from excessive current.

## Experiment 3: Resistance Per Unit Length and Resistivity of Nichrome Wire

This experiment has begun by connecting the digital multimeter (DMM) across varying lengths of a nichrome wire mounted along a meter scale. We have moved the negative probe of the DMM along the wire from the 0 cm mark toward the far end, while keeping the positive probe fixed at one end (point A). At each selected length, we have recorded the scale reading l in centimeters and the corresponding voltage V measured by the DMM. A graph has been plotted with length (l) on the x-axis and voltage (V) on the y-axis. The resulting straight line has indicated a uniform potential drop along the wire, and the slope of this graph has provided the voltage drop per unit length, found to be 0.033 V/cm.

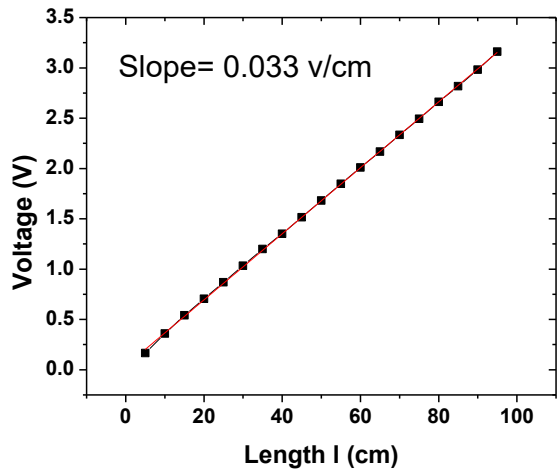

We have used the closed-circuit voltage across the full 96 cm nichrome wire, measured by the DMM as 3.306 V, and the total resistance of the wire, measured as 3.22 Ω, to calculate the current flowing through the wire: $I = \frac{V}{R} = 1.027\ A$. Using this current and the voltage drop per unit length; we have calculated the resistance per unit length is 0.0321Ω/cm. We have also measured the diameter of the wire using a screw gauge is 0.74mm and length 96 cm. So, the value of resistivity (ρ) resistivity of nichrome is $1.44 \times 10^{-4}$ Ω·cm. This experiment has strengthened our understanding of voltage measurement techniques, graph-based analysis, and the practical determination of resistivity, connecting experimental skills with theoretical concepts in electricity.

Fig 2: Voltage drop vs Length graph

Experiment 4: Verification of Ohm's Law

We have connected Mobile charger between two terminal A and B of the wire. Positive terminal is connected with A and negative with B. Point A has been also connected to the left terminal of the galvanometer through a 10K-ohm carbon resistor. The free end of a long wire connected to the right terminal of galvanometer has been allowed to move along the wire from point A so that the deflection of the galvanometer needle has remained within its scale. The current I in the resistor has been calculated using the product of the galvanometer's figure of merit (FOM) and the deflection d in microamperes. We have also recorded the corresponding lengths of the wire from point A to where the contact has been made. This setup has effectively acted as a potential divider circuit. By multiplying the length of the wire with the previously calculated voltage per unit cm, we have determined the corresponding voltage at each position. A graph of voltage V versus current I has been plotted with voltage along the x-axis and current along the y-axis. From the slope of this graph, we have determined the resistance of the galvanometer circuit.

Table 1: Data for Ohm's Law Verification

| Length of the connected terminals, l (cm) | Galvanometer Deflection, d | Voltage= l x length per unit length (V) | current= d x FOM (µA) |
|---|---|---|---|
| 10 | 4 | 0.33 | 68 |
| 20 | 7 | 0.66 | 119 |
| 30 | 11 | 0.99 | 187 |
| 40 | 14 | 1.32 | 238 |
| 50 | 18 | 1.65 | 306 |
| 60 | 22 | 1.98 | 374 |
| 70 | 26 | 2.31 | 442 |
| 80 | 30 | 2.64 | 510 |

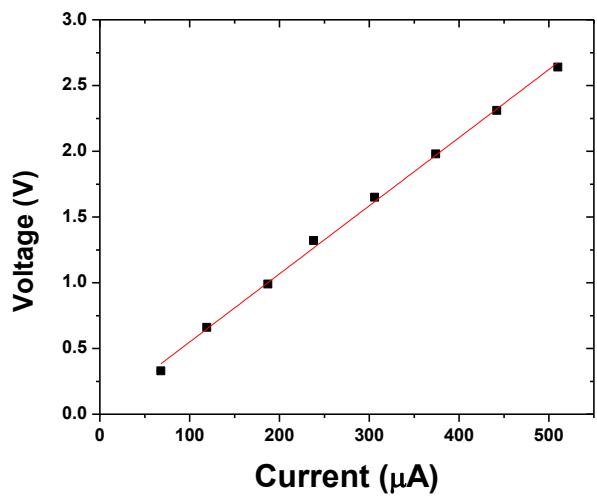

Fig 3: Ohm's Law                                                                                   verification Graph

Experiment 5: Characteristics of a non-Ohmic device

We have replaced the resistor with a Zener diode and have used the same setup to measure voltage and current. The voltage–current graph plotted from our measurements has revealed the threshold behaviour of the diode. This nonlinear relationship has introduced us to the fundamentals of semiconductor physics and the behaviour of nonlinear devices.

Fig 4: Characteristics of a non-Ohmic device

## Experiment 6: EMF of a Chemical Cell Using a Potentiometer Setup

In this experiment, the Indigenous Metre Bridge (IMB) has been used as a potentiometer. The power supply from the mobile charger has been connected across the full length of the bridge wire to create a uniform potential gradient. A chemical cell—such as a dry cell or a potato cell with zinc and copper electrodes—has been inserted into the circuit by placing it in the position normally occupied by a resistor or diode, as shown in the experimental setup. One terminal of the cell has been connected to point A on the IMB, while the other terminal has been connected through the galvanometer. The free end of a wire from the galvanometer has then been used to probe along the bridge wire until a point of zero deflection has been observed. This balancing length has corresponded to the point where the potential drop along the wire equals the EMF of the cell.

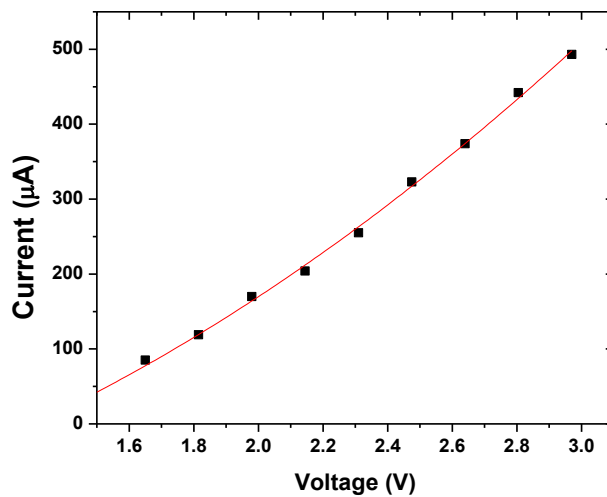

Using the relation: $e = V\left(\frac{l}{L}\right)$, where V is the potential across the full-length L of the wire and l is the balancing length, the EMF of the cell has been determined. In this case, with l=42 cm, L=96 cm, and V=3.306 V, the EMF of the dry cell has been calculated as 1.45V. This experiment has familiarized students with null-point techniques and has demonstrated the precision of potentiometric measurements using an extremely simple and accessible apparatus.

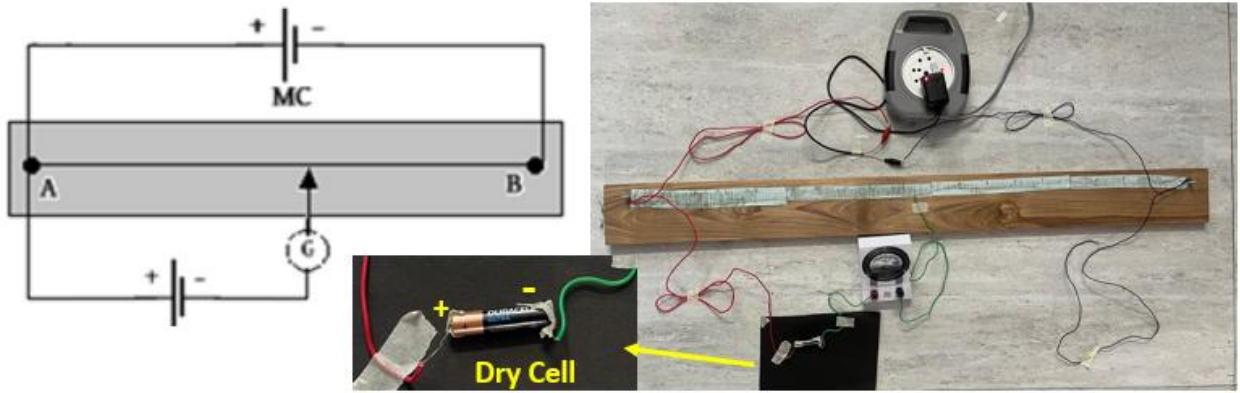

Fig 5: IMB set-up for deamination of EMF of a Chemical Cell

## Experiment 7: Determination of unknown Resistance Using the IMB as a Metre Bridge

In this experiment, we have used the Indigenous Metre Bridge (IMB) as a Wheatstone bridge to determine the resistance $R_X$ of an unknown carbon resistor. One arm has contained a known resistor R=10 Ω, and the other arm has held the unknown resistor $R_X$. We have connected the galvanometer between the midpoints of the two resistors, and have adjusted the sliding contact along the wire until no deflection has been observed, indicating a balanced condition.

With a total wire length of L=96 cm and a balancing length l=30.5 cm, we have calculated the unknown resistance as: $R_X = R \times \frac{L-l}{l} = 21.48 \Omega$.

By comparing the colour code, we have found the actual resistance to be 22 ohms. This experiment has allowed us to successfully demonstrate how the Indigenous Metre Bridge can be used to determine an unknown resistance using the principle of the Wheatstone bridge. The calculated resistance (21.48 Ω) has closely matched the colour-coded theoretical value (22 Ω), which has validated the accuracy of our method. Additionally, this activity has enhanced our skills in both practical circuit analysis and resistor identification through colour coding.

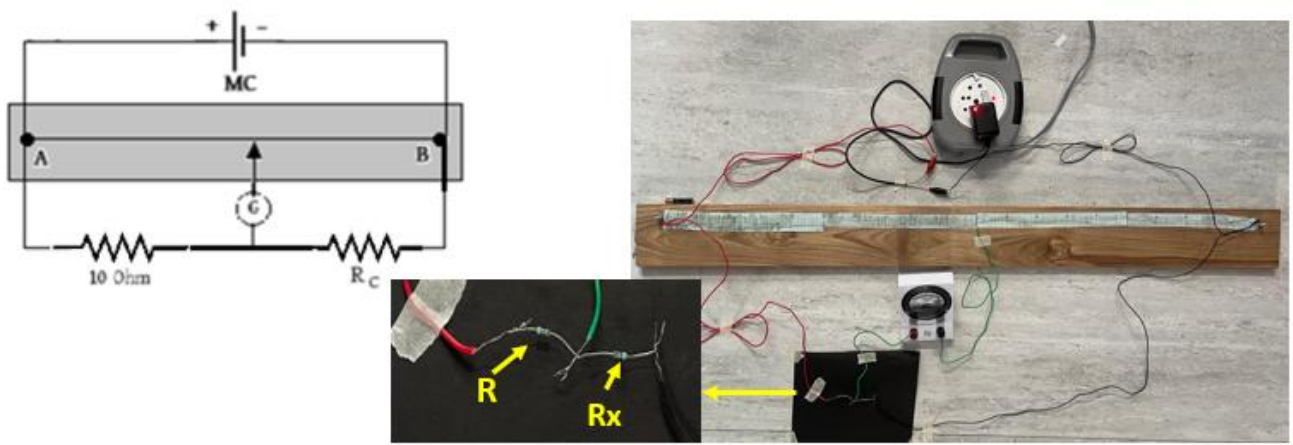

Fig 6: IMB set-up for determination of unknown resistance

Experiment 8: Series and Parallel Combinations of Resistors

In the final experiment, we have measured two individual carbon resistors of 22 Ω and 6.8 Ω, and then have combined them in both series and parallel configurations. We have verified the theoretical formulae for equivalent resistance using our experimental data, which has helped strengthen our understanding of circuit analysis and link theory with practical observations.

For the series combination, we have obtained an experimental equivalent resistance of approximately 28.5 Ω, which has closely matched the theoretical value calculated as: $R_S = R_1 + R_2 = 28.8$ Ω.

For the parallel combination, we have measured an experimental equivalent resistance of approximately 5.07 Ω, which has agreed well with the theoretical value calculated using: $\frac{1}{R_P} = \frac{1}{R_1} + \frac{1}{R_2} = 5.19$ Ω

These results have confirmed the validity of the formulas for resistors in series and parallel, and the experiment has provided a meaningful hands-on experience in analysing basic resistor networks.

## Conclusion

This device has gone beyond being a mere instructional tool—it has emerged as a practical solution to the distinct challenges faced in both rural and urban science education. To assess its classroom applicability, we conducted two three-day workshops in West Bengal: one at Anandapur High School, a rural institution in Paschim Medinipur, and another at Vidyasagar Vidyapith Girls' High School, an urban school in Midnapore town. A total of 59 students from the rural school and 75 from the urban school participated. These programs, organized by the Indian Association of Physics Teachers (IAPT), Regional Council 15, provided students with hands-on experience using the Indigenous Metre Bridge (IMB).

In both settings, students successfully completed all experimental activities and responded with great enthusiasm. Rural participants appreciated the accessibility of the setup, while urban students expressed increased confidence in engaging with lab work. The accuracy and reliability of this method have also been clearly demonstrated through consistent experimental results that closely matched theoretical values. These outcomes confirm the IMB's versatility and effectiveness across diverse educational contexts, reinforcing its potential for widespread adoption and its role in promoting inclusive, experiential learning.

## Acknowledgement

We gratefully acknowledge the Indian Association of Physics Teachers (IAPT), Regional Council 15, for organizing and supporting this initiative. We extend our sincere thanks to the Midnapore College Centre for Scientific Culture for providing instrumental support and resource persons, without which this program would not have been possible.

We are deeply indebted to the dedicated team of resource persons whose efforts and expertise greatly enriched the workshops: Mr. Kalyan Mukhopadhyay (IPS), Dr. Makhan Lal Nanda Goswami, Dr.

Rajsekhor Bor, Dr. Birendra Nath Das, Dr. Shinjinee Das Gupta, Dr. Manimala Das, Mr. Pradip Mahanta, Mr. Sukumar Bera, Mr. Anirban Samanta, Mr. Tanumoy Pal, Mr. Papun Mondal, Mr. Ramesh Samanta, Mrs. Amrita Maity, Dr. Sumana Bhadra, Mrs. Rita Nag, Mrs. Kakali Khan, Mrs. Ruma Dash, Mr. Suman Kumar Dash, Mr. Sandip Sarkar, Mr. Soumentdra Bera, Mr. Jmiur Rahaman, Mr. Uttam Majhi, Mr. Subhajit Roy, and Mr. Farukh Mallick.

Our heartfelt thanks go to the Headmasters of the participating schools—Mr. Bisweswar Mondal of Anandapur High School and Mrs. Swati Bandhopadhyaya of Vidyasagar Vidyapith Girls' High School—for their wholehearted cooperation and support.

We are also thankful to Prof. Satyaranjan Ghosh, Teacher-in-Charge of Midnapore College, for his encouragement and academic guidance throughout the program. Finally, we acknowledge the invaluable assistance of Mr. Samir Dhal and Miss Sabit Guin, whose support behind the scenes helped ensure the smooth execution of the workshops.

This collective effort has truly demonstrated the strength of collaboration in advancing science education at the grassroots level.

## References:

1. Z. Uddin, N. Sadiq, Simulating Physics Experiments in Spreadsheets II—Experimenting with the Half-Deflection Method, Resistance of a Voltmeter, and a Meter Bridge, Phys. Teach., Vol.63, No.4, pp.281–285, 2025.
2. N.D. Setyani, Suparmi, Sarwanto, J. Handhika, Students' conception and perception of simple electrical circuit, J. Phys.: Conf. Ser., Vol.909, 012051, 2017.
3. D. Shipstone, Pupils' understanding of simple electrical circuits. Some implications for instruction, Phys. Educ., Vol.23, No.2, pp.92, 1988.
4. L. Liégeois, E. Mullet, High school students' understanding of resistance in simple series electric circuits, Int. J. Sci. Educ., Vol.24, No.6, pp.551–564, 2002.